\documentstyle[12pt,fps]{article}

\def\lsi{\raise0.3ex\hbox{$<$\kern-0.75em\raise-1.1ex\hbox{$\sim$}}}
\def\gsi{\raise0.3ex\hbox{$>$\kern-0.75em\raise-1.1ex\hbox{$\sim$}}}
\newcommand{\lsim}{\mathop{\lsi}}
\newcommand{\gsim}{\mathop{\gsi}}

\newcommand{\R}{{\kern+.25em\sf{R}\kern-.78em\sf{I} 
\kern+.78em\kern-.25em}}

\begin{document}

\begin{flushright}
HU-EP-02/12 \\
\end{flushright}

\vspace*{1cm}

\begin{center}

{\LARGE\bf Convergence Rate and Locality }

\vspace*{5mm}

{\LARGE\bf of Improved Overlap Fermions}

\vspace{2cm}

W. Bietenholz

\vspace*{6mm}

Institut f\"{u}r Physik \\
Humboldt Universit\"{a}t zu Berlin \\
Invalidenstr. 110 \\ 
D-10115 Berlin, Germany \\

\end{center}

\vspace*{1.5cm}

{\it We construct new Ginsparg-Wilson fermions for QCD
by inserting an approximately chiral Dirac operator ---
which involves ingredients of a perfect action ---
into the overlap formula. This accelerates the convergence
of the overlap Dirac operator by a factor of 5 compared to the
standard construction, which inserts the
Wilson fermion as a point of departure. Taking into account the
effort for treating the improved fermion, we are left with an
total computational overhead of about a factor 3. This remaining
factor is likely to be compensated by other virtues; here we show
that the level of locality is clearly improved,
so that the exponent of the correlation decay is doubled.
We also show that approximate rotation invariance
is drastically improved, but a careful scaling test has to be
postponed.}

\newpage

\section{Introduction}

Over the recent years, substantial progress has been achieved in
the long-standing problem of constructing a formulation of chiral 
fermions on the lattice. It turned out that there is a
particularly harmless way to break the full chirality of 
the the lattice Dirac operator, so that the physical 
properties related to chirality are still represented correctly.
This breaking term is sufficient to circumvent the
Nielsen-Ninomiya No-Go theorem \cite{NN}, which refers to full 
chirality, in the sense that the lattice Dirac operator $D$
anti-commutes with $\gamma_{5}$. This harmless breaking 
is characterized by the {\em Ginsparg-Wilson relation} (GWR) 
\cite{GW}
\begin{equation} \label{GWR}
\{ D , \gamma_{5} \}_{x,y} = 2 (D \gamma_{5} R D)_{x,y} \ ,
\end{equation}
where the kernel $R$ has to be {\em local} and it must not anti-commute
with $\gamma_{5}$. The standard choice for this term reads
\begin{equation} \label{Rst} 
R^{st}_{x,y} = \frac{1}{2\mu} \delta_{x,y} \ , \qquad
\mu > 0 \ .
\end{equation}

The general class of solutions to the GWR with the standard kernel 
$R^{st}$ is given by the so-called {\em overlap formula} \cite{Neu}
\begin{eqnarray}
D_{ov} &=& A_{ov} + \mu \nonumber \\
A_{ov} &=& \mu A_{0} / \sqrt{A_{0}^{\dagger} A_{0}} \nonumber \\
A_{0} &=& D_{0} - \mu \ . \label{overlap}
\end{eqnarray}
For $D_{0}$ one may insert some sensible lattice Dirac
operator. It has to be local (in the sense that the
correlations decay exponentially) and free of
fermion doubling, but we do not require any form of chirality
for $D_{0}$. The resulting operator $A_{0}$ is then transformed
into an operator $A_{ov}$, so that $\frac{1}{\mu}A_{ov}$
is {\em unitary}. This provides the overlap 
Dirac operator $D_{ov}$, which solves the GWR.
The choice of the mass parameter $\mu$ is constrained;
inside its allowed interval it can be optimized with 
respect to certain criteria, see below.

From eq.\ (\ref{overlap}) it is obvious
that the spectrum of a GW fermion --- with the standard kernel
given in eq.\ (\ref{Rst}) --- is always situated on a circle
in the complex plane with center and radius $\mu$; we denote
it as the {\em GW circle}. This condition for the spectrum is 
equivalent to the GWR with kernel $R^{st}$.

In almost all the literature, the Wilson operator
$D_{W}$ was inserted as $D_{0}$ into the overlap formula, without
consideration of other options. We denote the resulting 
particular overlap fermion as the {\em Neuberger fermion}. As any
Ginsparg-Wilson fermion (GW fermion), it has a chiral symmetry 
which is lattice modified but exact \cite{ML}. 
Since the corresponding transformation
is local --- based on the locality of $R$ --- the chiral limit
is not plagued by mass renormalization.

The generalization of the overlap fermions to a whole class
of solutions of the GWR --- and the motivation for considering alternative
operators $D_{0} \neq D_{W}$ --- were given in Ref.\ \cite{EPJC}.
In fact, there are further issues of importance for a lattice
fermion formulation, such as the quality of the {\em scaling} behavior,
the level of {\em locality} and of approximate {\em rotation} invariance.
With these respects, the Neuberger fermion is not quite satisfactory,
even though it is local up to moderate coupling strength \cite{HJL}.
All these properties depend on the choice of $D_{0}$, hence
it is motivated to search for a better starting
operator than $D_{W}$.

Moreover, the practical evaluation of $D_{ov}$ is a formidable
numeric task. In particular, the quenched simulation of
the Neuberger fermion is about as expensive as the simulation
of dynamical Wilson fermions \cite{HJLe}, i.e.\ the computational
overhead amounts to about two orders of magnitude.
Also with this respect a better choice of $D_{0}$ can help.
Our guide-line is the following observation:
if we insert a GW fermion (with $R^{st}$) as $D_{0}$, then
it is simply reproduced by the overlap formula (\ref{overlap}),
$D_{ov}=D_{0}$.
\footnote{Of course, this property holds for any fixed GW kernel $R$,
if one uses the suitably generalized overlap formula \cite{EPJC}.}
Therefore, in practice we try to insert an approximate GW fermion
as $D_{0}$ \cite{EPJC}, so that it is altered only a little by the 
overlap formula,
\begin{equation} \label{approx}
D_{ov} \approx D_{0} \ .
\end{equation}
This small alteration corresponds to a {\em chiral correction}
(in the sense of the GWR).
\footnote{Similarly, one could also insert an approximately chiral 
$D_{0}$ in the 4d space of domain wall fermions \cite{EPJC,Shamir}.}

Thus the first issue is to construct an approximate GW fermion 
``by hand''. At this point, we recall that also perfect \cite{GW}
and classically perfect fermions \cite{HLN} solve the GWR. Their
construction is very tedious, but in the present context we
can use a relatively simple approximation, namely a {\em hypercube
fermion} (HF). The mass renormalization is still quite strong
for simple HFs \cite{BBCW,precon}, which is a problem in their 
direct application \cite{OBBCW}.
However, here we reach the chiral limit by inserting the
HF into the overlap formula.

Of course, the (classically) perfect fermion has further virtues
in addition to chirality. In particular its scaling behavior is (almost)
free of lattice artifacts, i.e.\ dimensionless ratios of
physical observables are (almost) independent of the lattice
spacing. Moreover the observables are also rotational invariant
for perfect fermions, and rotation symmetry is approximated 
very well for the truncated perfect hypercube fermions,
as the pion dispersion shows \cite{BBCW}.
Since the overlap formula modifies the HF just a little --- see
relation (\ref{approx}) --- we expect the good scaling and rotation 
behavior to persist under the chiral correction. Then we obtain
an {\em improved overlap fermion}.

There is yet another virtue to be expected based on relation
(\ref{approx}): the hypercube fermion is short-ranged ---
its free couplings are restricted to a unit hypercube on 
the lattice --- hence we also expect a {\em high level of locality}
for the overlap-HF (given by $D_{ov}$ if we insert $D_{0}=D_{HF}$).
Due to the modest alteration, the long range couplings can
be turned on just slightly, hence their exponential
decay will be fast. The Wilson fermion is also short-ranged,
but it changes a lot in the transition to the Neuberger
fermion, so we do not have the above reason to expect a good
locality. Indeed, even for the free Neuberger fermion
the decay of couplings (in the separation of fermion and
anti-fermion) is a rather slow exponential \cite{EPJC}.

{\em All} these expected virtues of the overlap-HF have been 
tested and impressively confirmed in the 2-flavor
Schwinger model \cite{BH}.
Now we carry on this program to QCD. In Section 2, we first 
describe the construction of a suitable HF in $d=4$, and we 
illustrate its proximity to a GW fermion by evaluating typical
fermionic spectra on small lattices.
In Section 3 we discuss the polynomial evaluation of the inverse
square root in eq.\ (\ref{overlap}). In Section 4
we investigate the speed of the numeric evaluation
of the overlap formula. We show that for our 
HF a polynomial of a low order is sufficient to
approximate the sign function or the inverse square root to a 
high accuracy. For the Neuberger fermion the same accuracy 
requires a polynomial of a much higher degree.
Section 5 gives results for the condition numbers
and their impact on the convergence rate, now proceeding
to larger lattices. 
Section 6 presents a comparative study of the level of
locality, i.e.\ of the exponential decay of the maximal 
correlation at large distances, and Section 7 compares
the violations of rotation invariance.
Section 8 contains a summary and our conclusions.

\section{Approximate Ginsparg-Wilson fermions in QCD}

For free or perturbatively interacting fermions, 
perfect actions can be constructed analytically \cite{QuaGlu}. 
In the case of non-perturbative interactions, this is possible 
only numerically and only in the classical approximation.
\footnote{For recent work on perfect actions for semi-classical
effective actions, see Ref.\ \cite{monopol}.}
The renormalization group transformation
leading to the (quantum) perfect action would involve
effectively a continuum functional integral. Still its
existence is of conceptual interest. It implies
for instance that also a topological lattice charge without
lattice artifacts exists \cite{topo}, and that supersymmetry can in 
principle be represented continuously on the lattice \cite{susy}.
Similarly, the perturbatively perfect action shows that
e.g.\ the axial anomaly is reproduced correctly \cite{BWPLB}, 
but only a modest improvement persists
on the non-perturbative level \cite{OBBCW,BS}.
What is (in principle) accessible and promising for simulations is
an approximation to the classically perfect action \cite{HN}, 
which works well for fermions in two dimensions \cite{LP}. However,
the difficult construction and application  in $d=4$ is still 
under investigation \cite{Bern}. In that case, there is also a
potential for an excellent scaling, but in order to obtain
a sufficient chiral quality, it seems that the group working on
it (the Bern collaboration) also depends on the concept of 
Refs.\ \cite{EPJC,BH} (chiral correction).

In the free case, the truncated perfect HF has still
an excellent scaling behavior \cite{BBCW,thermo}, if the 
renormalization group parameters are optimized for locality. 
Hence we use the free HF as the point of departure in our
attempt to construct a HF, which is an approximate GW 
fermion, and which is also promising with respect to 
scaling and approximate rotation invariance.
A few elements for its gauging are then added such
that the GWR is violated only modestly at the coupling
strength of interest. In this Section, we are going to 
describe this construction step by step. A synopsis of
this construction has been anticipated in Refs.\ 
\cite{GuangBang}. 

The concept formulated in Refs.\ \cite{EPJC,BH} has also been
adapted in Ref.\ \cite{TDG}, where some progress is reported,
although a very simple (``planar'') operator $D_{0}$ was used, 
which is quite far from a GW fermion even in the free case.
With this ``planar overlap operator'' some results for
the finite size scaling of the chiral condensate of 
Ref.\ \cite{HJLl} were reproduced, and the effects of instantons
in the chiral symmetry breaking were reconsidered \cite{TdGAH}.
Other very simple non-standard operators $D_{0}$ were used
in Refs.\ \cite{AB,Austra}. In contrast,
a very complicated approximate 
Ginsparg-Wilson fermion was constructed
in Ref.\ \cite{Gatt} by introducing many parameters and
tuning them for a minimal GWR violation at a specific value of $\beta$.
This corresponds to the first part of our program (constructing
an approximate GW fermion by hand), but since the overlap
formula is not used, chirality is still not exact. Moreover,
there are no ingredients in favor of improving other properties.
However, that work reports some gain if a specific improved
gauge action is used. Different improved gauge actions
were applied in Refs.\ \cite{Bern,DLLZ}. 
The use of improved gauge actions is also on our agenda, but in the
present work we always use the standard plaquette gauge action.
This allows us to observe unambiguously the progress due to the
improved Dirac operator. 

All the considerations below are based on quenched
configurations on periodic lattices of sizes $4^{4}$,
$8^{4}$ (Sections 2, 3, 4) and $12^{4}$ (Sections 5, 6 and 7), 
and beyond Section 2 we always use $\beta =6$. \\

{\bf Step 1: Minimal gauging}\\

As we mentioned above, we start from the couplings of the
free perfect fermion. It is optimized for locality, and then
truncated to a HF by imposing periodic boundary conditions.
We write the resulting Dirac operator as
\begin{equation}
D(x-y) = \rho_{\mu}(x-y) \gamma_{\mu} + \lambda (x-y) \ ,
\qquad (x,y \in Z\!\!\!Z ^{4}, 
\ \vert x_{\nu}-y_{\nu} \vert \leq 1 )
\end{equation}
and the couplings of the vector term $\rho_{\mu}$ and
the scalar term $\lambda$ are given in Ref.\ \cite{BBCW},
Table 1. Note that the components of $x-y$ are restricted
to $-1,0,1$, and that $\rho_{\mu}$ is odd in the $\mu$-direction
and even otherwise, while $\lambda$ is entirely even.
In the free case, this HF is an excellent approximation
to a Ginsparg-Wilson fermion \cite{EPJC}.

Our first HF for QCD, with a Dirac operator of the form
\begin{equation}
D(x,y,U) = \rho_{\mu}(x,y,U) \gamma_{\mu} + \lambda (x,y,U) \ ,
\qquad (U \ : \ {\rm compact~gauge~field})
\end{equation}
is now obtained by ``minimal gauging''
of the free HF: the sites, which are coupled in the HF
formulation, are connected by shortest lattice path only.
The gauging is done by simply attaching the free coupling to
these paths, divided into equal parts where several shortest 
paths exist.

\begin{figure}[hbt]
\hspace*{3cm}
\def\fpsangle{270}
\epsfxsize=80mm
\fpsbox{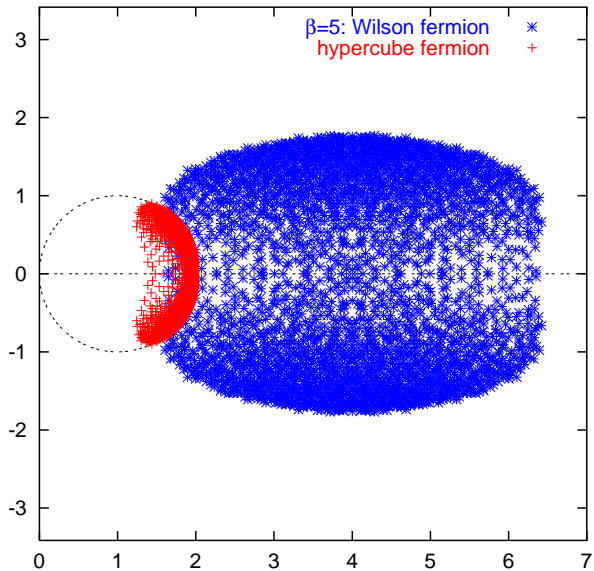}
\vspace{-3mm}
\caption{\it{The fermionic spectrum for the minimally gauged HF
and for the Wilson fermion, in a typical configuration at
$\beta =5$ on a $4^4$ lattice.}}
\label{beta5}
\end{figure}

This simple prescription already provides an excellent pion 
dispersion relation \cite{BBCW}. On the other hand, with
such a gauging the HF suffers from a strong additive mass 
renormalization \cite{BBCW,precon}, comparable to the Wilson
fermion. We illustrate this at strong coupling in 
Fig.\ \ref{beta5}: it shows the full Dirac operator spectra for 
the Wilson fermion and the HF for a typical configuration at 
$\beta =5$ on a $4^{4}$ lattice.

In such a rough gauge background, the overlap formula
is only applicable if we have an excellent GW approximation
to be inserted as $D_{0}$. Fig.\ 1 shows that the standard
Neuberger fermion, $D_{0}=D_{W}$, certainly fails, and the 
minimally gauged HF cannot handle such a strong coupling either
\cite{Dubna}.
\footnote{Note that in such a situation it could be misguiding
to consider only the spectrum of $A^{\dagger}_{0}A_{0}$, as it 
is sometimes done in the literature. For the HF in Fig.\ \ref{beta5}, 
that spectrum alone would look quite satisfactory, even though the 
physically crucial left arc is missing.}

Since we assume the kernel $R$ to be of the form (\ref{Rst}),
the overlap formula performs a sort of projection of the
eigenvalues onto the GW circle. This projection
tends to be close to radial \cite{Graz-spec,BH}.
Hence we need a statistically safe distinction between a
left arc and a right arc of the spectrum of $D_{0}$ to start with,
so that it can be mapped onto a GW circle in a reliable way.
Here we focus on the regime close to the real axis. Eigenvalues have
to cross the center as one changes topological sectors
--- since we define the topological charge by the index theorem 
\footnote{This is a sensible definition of the topological
charge on the lattice for any Ginsparg-Wilson
fermion \cite{ML}. Up to moderate coupling strength, it is
often close to the geometrical charge 
\cite{Graz-topo}. In general it also tends to agree with
the charge identified from cooling the $SU(N)$ gauge
configurations, especially at increasing $N$ \cite{Wenger}.
In the case of classically perfect fermions
it corresponds to the classically perfect charge \cite{HLN},
and analogously for (quantum) perfect fermions.}
--- but they have to do so quickly as the deformation proceeds, so 
that the distinction is clear for the typical configurations at the 
coupling strength under consideration.

However, in the present case the freedom of choosing this center 
does still not help: wherever we choose $\mu$, it will frequently
happen that too many zeros occur (due to too many mappings to
the left), so that the doubling problem is back, or that too
many eigenvalues are mapped to the right-hand side, so that mass
renormalization re-appears. Also further generalizations of $R$
(beyond the restriction to one site) do not help in this situation.

\begin{figure}[hbt]
\begin{tabular}{cc}
\def\fpsangle{270}
\epsfxsize=60mm
\fpsbox{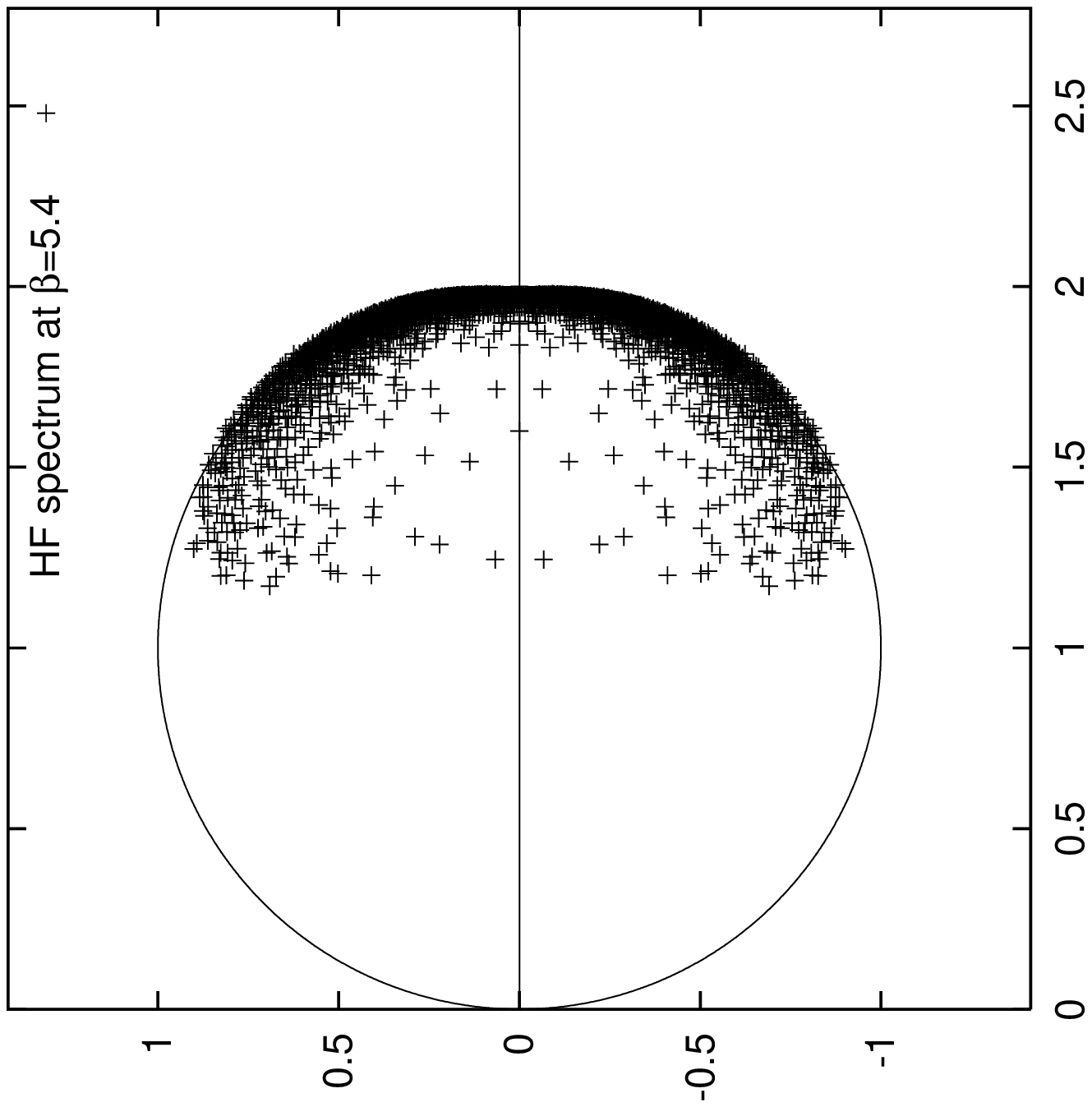}
\epsfxsize=60mm
\fpsbox{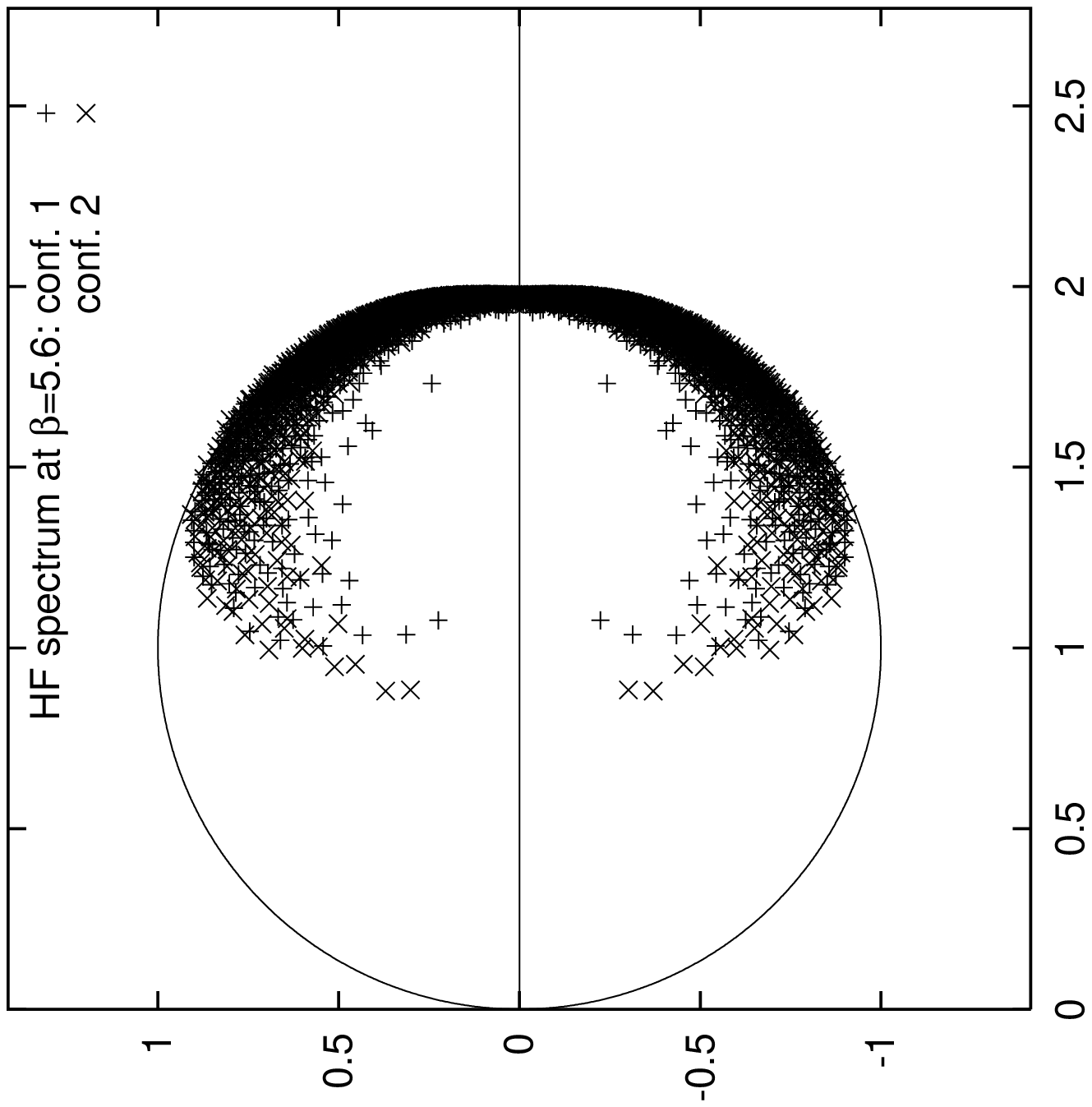}
\end{tabular}
\vspace{-3mm}
\caption{\it{Typical spectra of the minimally gauged HF at
$\beta = 5.4$ (on the left) and at $\beta = 5.6$ (on the right). 
In the latter case, the overlap formula seems to be applicable 
with mass parameter $\mu \approx 1.4$.}}
\label{b5.4b5.6}
\end{figure}

A question of interest is at which coupling strength
the applicability of the overlap formula with simple operators
like $D_{HF}$ sets in. Typical spectra on a $4^{4}$ lattice
suggest that for instance 
at $\beta =5.4$ the coupling is still too strong, 
but at $\beta \gsim 5.6$ we are about to approach to safer 
grounds, see Fig.\ \ref{b5.4b5.6}. 
The same is true for the Wilson fermion
(see first Ref.\ in \cite{GuangBang}, Fig.\ 8). On larger 
lattices the minimal $\beta$ is still likely to
rise to about $\beta \gsim 5.7$.\\

{\bf Step 2: Critical link amplification} \\

Worried about the strong mass renormalization, we first want
to move our HF towards the chiral limit. We do so by amplifying
each link variable by a factor $1/u$, 
\begin{equation}
U_{\mu}(x) \ \rightarrow \ \frac{1}{u} \, U_{\mu}(x) \ , \qquad
u \lsim 1 \ .
\end{equation}
The idea is to compensate the (mean) link suppression due to
the gauge field. This can be viewed as a generalization of the 
critical hopping parameter used for Wilson fermions, but it is
also related to the spirit of ``tadpole improvement'' \cite{tad}.

\begin{figure}[hbt]
\hspace*{3cm}
\def\fpsangle{270}
\epsfxsize=80mm
\fpsbox{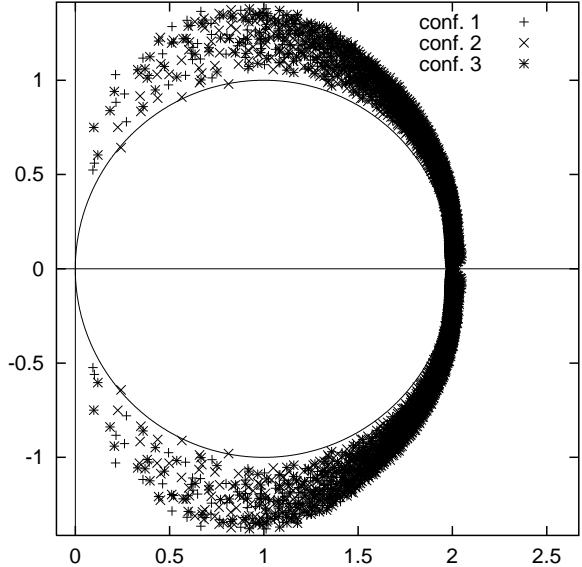}
\vspace{-3mm}
\caption{\it{HF spectra at $\beta =6$ on a $4^{4}$ lattice
at critical link amplification ($u=0.8$) for three configurations.}}
\label{ucrit}
\end{figure}
Fig.\ \ref{ucrit} illustrates that criticality is reached with
$u \simeq 0.8$ at $\beta =6$.
\footnote{The absence of the arc very close to 0 is due to the
{\em small size} of the lattice; it is {\em not} due to the
choice of the lattice Dirac operator.}
We see that this already provides
a decent approximation to a GW fermion. This is remarkable,
because we approach the chiral limit in the most economic way,
by staying within the framework of minimal gauging. We did not
introduce additional lattice paths yet, hence Step 2 does not
require any computational effort (once the critical value of 
$u$ is determined).\\

{\bf Step 3: Fat links} \\

We now want to improve the chiral quality of our HF further by
going beyond minimal gauging. As a non-minimal element we
introduce {\em fat links}. For a given configurations, we substitute 
each link variable according to the following scheme
\begin{equation}  \label{fatlink}
link \ \rightarrow (1-\alpha) \ link \ + \ \frac{\alpha}{6}
\Big[ \ \sum staples \ \Big] \ , \qquad \alpha \in \R \ ,
\end{equation}
{\em before} evaluating the Dirac operator. 
This is still an economic tool.
The substituted link variable on the
right-hand side is not mapped back onto the gauge
group, in contrast to the APE blocking \cite{APE},
and we do not iterate the substitution (\ref{fatlink}).

As a first observation, we note that the mass renormalization
is {\em enhanced} for increasing $\alpha$. This may seem 
counter-intuitive if one imagines that a positive $\alpha$
makes the configurations ``appear smoother''.
However, a more precise picture confirms this observation:
the only coupling of range 0 is $\lambda (0,0,0,0)=1.853\dots$
The rest of the scalar term is negative, 
and in the free case
we have $\sum_{x} \lambda (x) = 0$. The gauge field now suppresses
the negative contributions, whereas $\lambda (0,0,0,0)$ remains
unchanged, so that a positive mass sets in (if we keep $u=1$).
This is already true for minimal gauging, but if we add fat links
with $\alpha > 0$, this suppression of the negative part gets even
stronger. A fraction of the negative couplings is now attached to
staples instead of single links, and the staple suppression
corresponds to the third power of the mean link suppression.

\begin{figure}[hbt]
\begin{tabular}{cc}
\def\fpsangle{270}
\epsfxsize=60mm
\fpsbox{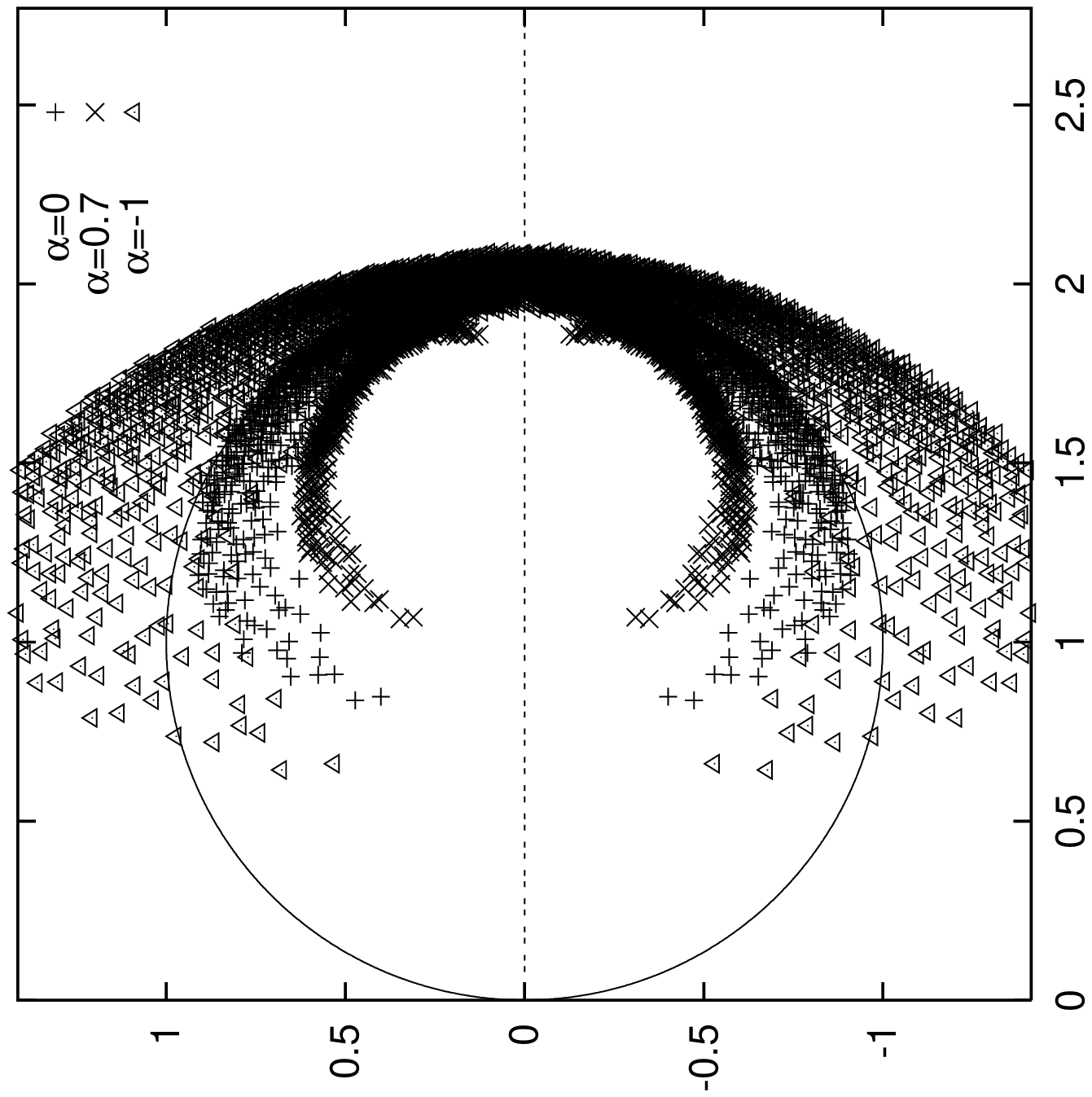}
\epsfxsize=60mm
\fpsbox{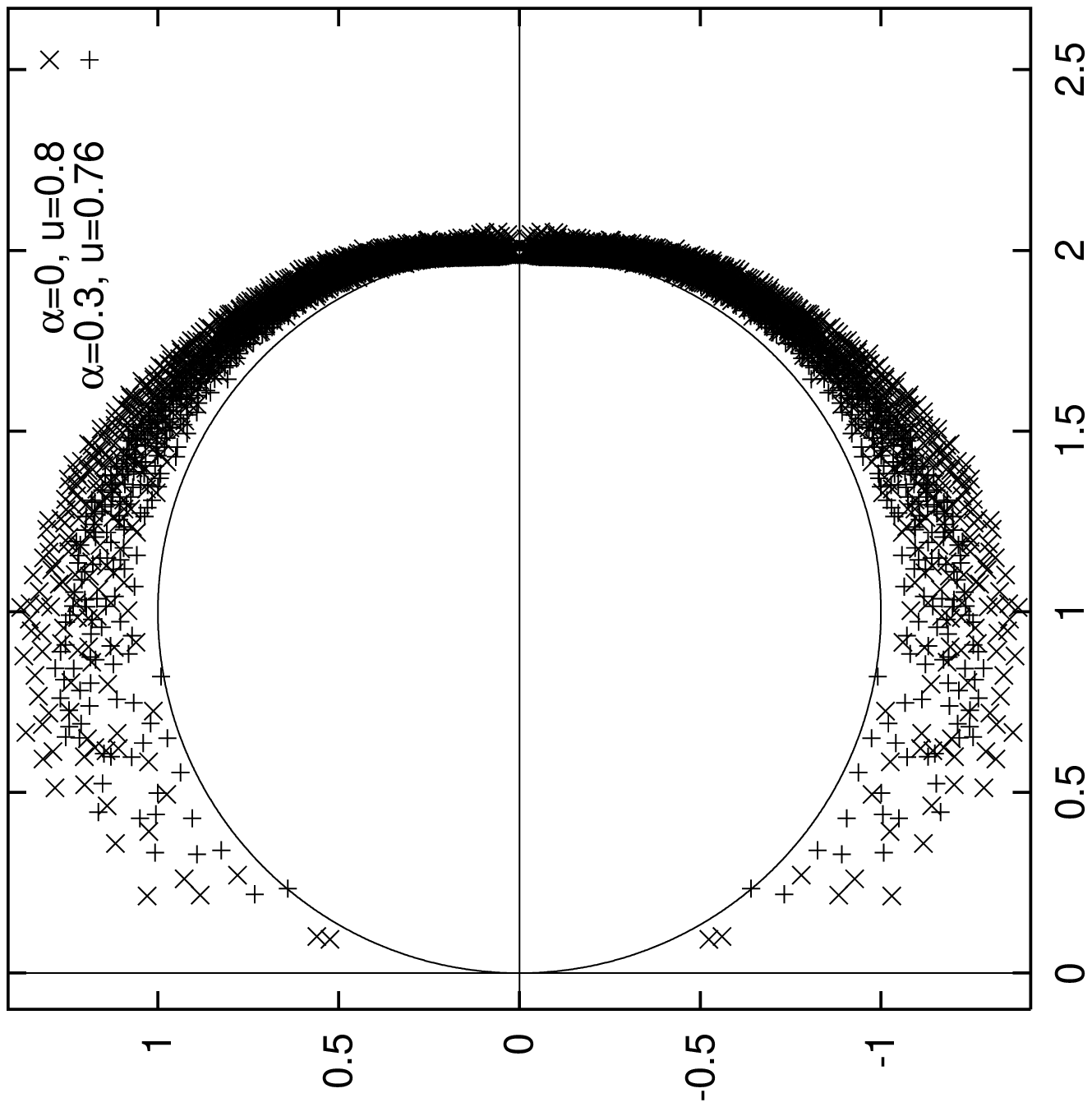}
\end{tabular}
\vspace{-3mm}
\caption{\it{The effect due to the variation of staple
coefficient $\alpha$ in the uncritical HF (on the left).
On the right: the spectrum of the critical HF with
and without fat links (everything at $L=4$, $\beta=6$).}}
\label{alf-fig}
\end{figure}

If we wanted to use fat links to move closer to the
chiral limit, we had to take $\alpha < 0$. Then the critical
value of $u$ rises, and for some strongly negative $\alpha$
(far below $-1$)
it even arrives at 1, so that criticality could, in principle, 
be realized solely by means of fat links.
However, in this case the rest of the
spectrum is very far from a GW circle --- in particular the
upper and lower arc are far outside the GW circle --- so we do not
recommend negative values of $\alpha$.
Fig.\ \ref{alf-fig} (on the left) shows this effect, and we also
see that positive $\alpha$ are more adequate to make the shape
of the spectrum circle-like.
So what is really favorable to improve the proximity to a GW 
fermion is a positive $\alpha$ along with an adapted (i.e.\ decreased)
critical value of $u$. A good choice is $\alpha = 0.3$, which
requires $u=0.76$ at $\beta =6$.
This reduces the radial fluctuation of the eigenvalues --- the
eigenvalues move closer together --- and the upper and lower arc
move closer to the GW circle. This is shown in Fig.\ 
\ref{alf-fig} (on the right), which compares the spectra with and 
without fat links for the same configuration.\\

{\bf Step 4: Vector term suppression} \\

The above picture for the mass renormalization ignores 
the r\^{o}le of the vector
term $\rho_{\mu}(x,y,U)$. In fact, some tests confirmed that it
has only a very modest influence on the location of the arc on
the left-hand side (and also on the right-hand side).
This location is essentially determined by the scalar
term, but the vector term is crucial for the height of the
upper and lower arc, i.e.\ the term $\rho_{\mu} \gamma_{\mu}$
is responsible for the imaginary part of the spectrum.

\begin{figure}[hbt]
\def\fpsangle{270}
\epsfxsize=110mm
\fpsbox{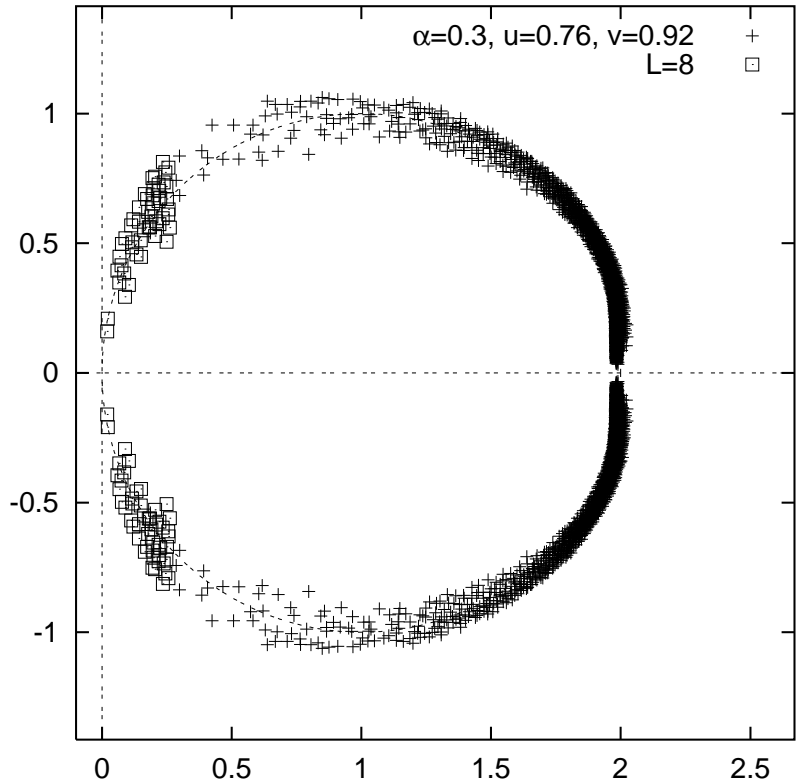}
\vspace{-3mm}
\caption{\it{The HF on a $4^{4}$ lattice with critical link
amplification, fat links and a suppression of the 
vector term. We also show the ``continuation'' around $0$
on a $8^{4}$ lattice with the same parameters, i.e.
$u=0.76$, $\alpha =0.3$, $v=0.92$.}}
\label{non+ultra}
\end{figure}

This property will now be used for a further improvement, still
without extra computational effort. We introduce different
link amplification factors for the vector term and the
scalar term, since they play a different r\^{o}le.
We first keep the critical factor $1/u$ as an overall link
amplification, but then we multiply a link suppression factor
$v$ only in the vector term. So now we multiply the links as follows:
\begin{eqnarray}
U_{\mu} \ \rightarrow \ \frac{v}{u} \ U_{\mu} && \quad {\rm in~~ } 
\rho_{\mu} (x,y,U) \nonumber \\
U_{\mu} \ \rightarrow \ \frac{1}{u} \ U_{\mu} && \quad {\rm in~~ } 
\lambda (x,y,U) \ ,
\end{eqnarray}
where $u \lsim v \lsim 1$.
Still the fat links are useful to suppress the radial fluctuations
of the eigenvalues, so we stay with $\alpha = 0.3$, $u=0.76$,
and the suitable vector term suppression amounts to
$v = 0.92$. Now the upper and lower arc also follow the GW
circle, and we obtain therefore a very satisfactory spectrum,
see Fig.\ \ref{non+ultra}.
It shows again the spectrum on a $4^{4}$ lattice, 
but this time we also include part of the spectrum of a typical
configuration on a $8^{4}$ lattice. From there we show the
100 eigenvalues with the smallest real parts, thus we also
visualize how the spectrum ``continues'' around zero.\\

{\bf A comment on the clover term} \\

\begin{figure}[hbt]
\hspace*{3cm}
\def\fpsangle{270}
\epsfxsize=80mm
\fpsbox{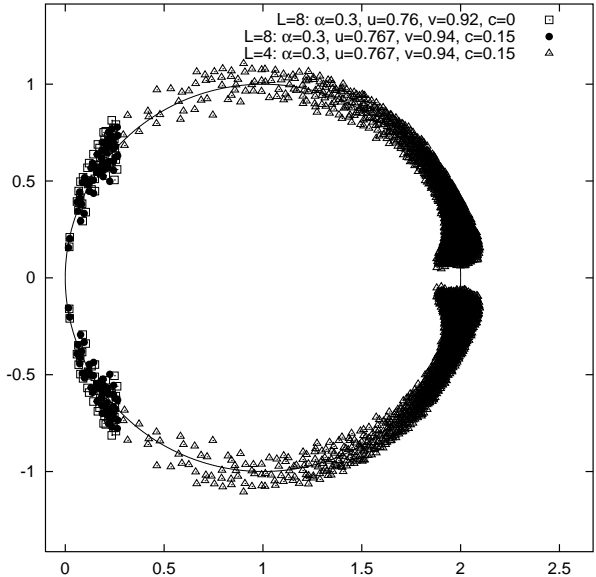}
\caption{\it{The effect of the clover term in the HF
spectrum.}} 
\vspace*{-3mm}
\label{clov-fig}
\end{figure}

An obvious candidate for a next step beyond minimal gauging
is the {\em clover term}. We performed a sequence of tests
with it being added to the HF versions discussed above.
We varied the clover coefficient and also considered both signs,
but from spectra on $4^{4}$ lattices 
we did not arrive at a clear further improvement
of the GW approximation in this way. A positive coefficient
has generally the effect to pull the spectrum closer
to the real axis, as it was observed before for the
Wilson fermion \cite{clover}, hence the optimal parameter
$v$ rises a little. A positive clover coefficient does
improve the arc around zero --- which appears on the $8^4$ 
lattice --- a little bit, but it distorts the rest.
As an example, we show the effect of the clover coefficient
$0.15$ in Fig.\ \ref{clov-fig}.
One could argue that it is precisely the left arc
which is physically crucial. However, we are going to insert
our HF into the overlap formula, so we end up with an
exact GW fermion anyhow. In view of the convergence rate
in the iterative evaluation of this formula, the maximal
deviation of the HF spectrum from the GW circle matters, so
one should not limit the attention to the left arc only.

Indeed, the systematic study of the transition to the
overlap fermion on larger lattices --- which will be presented 
in Section 5 --- shows
that a clover term with a small coefficient may help a little
to speed up the convergence. So we are going to include it 
on the $12^{4}$ lattice (used in Sections 5, 6 and 7), 
since it is computationally cheap anyhow.

\section{Polynomial approximations of the overlap formula}

In four dimensions, the evaluation of the overlap formula
is a notorious numeric problem. It can only be done by some 
approximation of the non-analytic function involved.
In particular, one tries to approximate it by a polynomial 
in the relevant interval. Here we apply Chebyshev polynomials
for that purpose, as suggested in Ref.\ \cite{HJL},
and we consider two possible ways to apply them.
For a systematic comparison of various approximation methods
in the case of the Neuberger fermion, see Ref.\ \cite{Wupp}.
We remark that from our experience, different types of
suitable polynomials do not lead to a very different
quality of the approximation (at fixed polynomial degree).

\subsection{Approximations of the sign function}

We first consider approximations of the sign function,
which is done in one way or the other in most of the literature. 
One writes the overlap formula as
\begin{eqnarray}
D_{ov} &=& \mu \Big[ 1 + \gamma_{5} \frac{H}{\sqrt{H^{2}}} \Big]
= \mu \ [ \, 1 + \gamma_{5} \epsilon (H) \, ] \nonumber \\
H &:=& \gamma_{5}(D_{0}-\mu ) = H^{\dagger} \label{signH}
\end{eqnarray}
and approximates the {\em sign function} 
\begin{equation}
\epsilon (x) = \left\{ \begin{array}{cccc} ~~ 1 &&& x>0 \\ 
-1 &&& x < 0
\end{array} \right.
\end{equation}
by a polynomial. As an alternative to the Chebyshev polynomials
that we are going to use here, also an ``optimal rational approximation''
has been suggested \cite{optrat} and it was applied for instance
in Refs.\ \cite{SCRI}.

The eigenvalue distribution of $H$ determines the interval
which is relevant for the approximation. Since this Section
is only meant to illustrate what the convergence rate depends on, 
we consider $\mu =1$ for simplicity. (Later, when we
study the convergence rate in detail, optimized mass parameters
$\mu$ will be inserted.)

\begin{figure}[hbt]
\begin{tabular}{cc}
\def\fpsangle{0}
\epsfxsize=60mm
\fpsbox{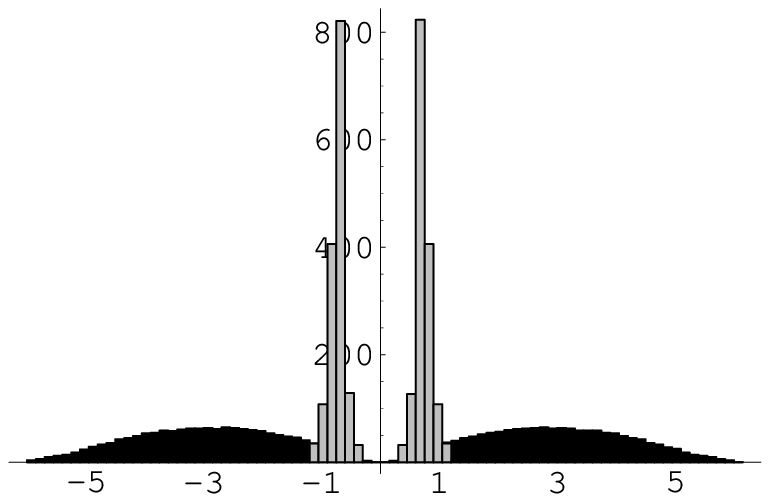}
\hspace{10mm}
\epsfxsize=60mm
\fpsbox{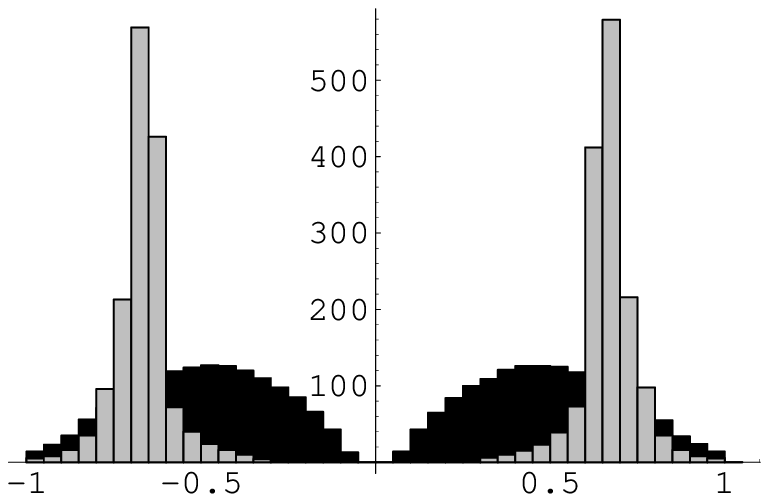}
\end{tabular}
\vspace{-3mm}
\caption{\it{On the left: the eigenvalue histograms for typical 
spectra of $H_{HF}$ (grey) and $H_{W}$ (black) at $\beta =6$. 
On the right: the same after re-scaling so that all 
eigenvalues have absolute values $\leq 1$.}}
\label{histoH}
\end{figure}

In Fig.\ \ref{histoH} (on the left) we show the eigenvalue 
histograms for the cases of the Wilson fermion, 
$H_{W} = \gamma_{5} (D_{W}-1)$, and of our preferred
HF on small lattices ($\alpha =0.3$, $u=0.76$, $v=0.92$), still at
$\beta =6$, on a $4^{4}$ lattice. 
We see that the spectrum of $H_{HF}$
is already sharply peaked at $\pm 1$, whereas for $H_{W}$
the distribution is very broad.

In order to make the polynomial approximation directly applicable 
for all eigenvalues, we first have to re-scale the spectra
so that they are entirely confined to the interval $[-1,1]$,
see for instance Ref.\ \cite{NR}.
The outcome of the minimal re-scaling
(division by the largest absolute value of an eigenvalue)
is shown in 
Fig.\ \ref{histoH} (on the right). We see that the spectrum of
$H_{HF}$ is not affected too much: the peaks are moved
to about $\pm 0.7$, but there is still a large gap around 
zero. This gap is important, because any polynomial 
approximation of the sign function is plagued 
by its worst errors near the discontinuity at zero
(remember for instance the notorious ``Gibbs phenomenon''
of the Fourier expansion).
On the other hand, for $H_{W}$
there is (after re-scaling) a considerable eigenvalue 
density around zero. This shows that it requires much more
effort to transform the Wilson fermion into an overlap fermion.

\begin{figure}[hbt]
\hspace*{3cm}
\def\fpsangle{270}
\epsfxsize=80mm
\fpsbox{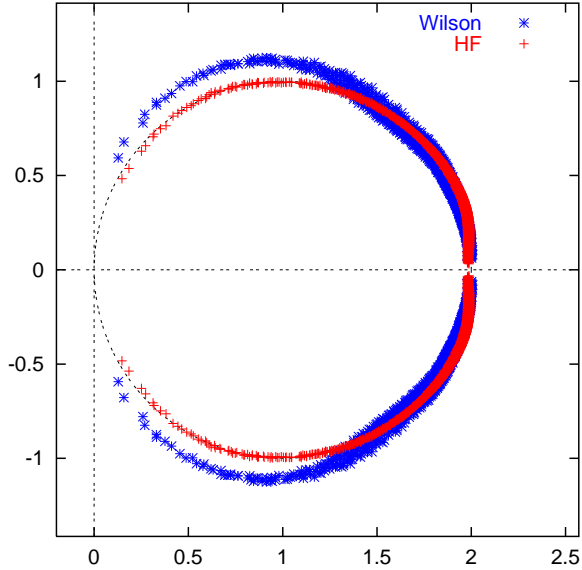}
\vspace{-3mm}
\caption{\it{The spectra of approximate overlap fermions
on a $4^{4}$ lattice, where the sign function is approximated 
by a polynomial of degree 21. We use a
configuration typical at $\beta =6$ and start from 
$D_{0}=D_{W}$ (stars) resp.\ from $D_{0}=D_{HF}$ (crosses).}}
\label{sign-poly}
\vspace*{-3mm}
\end{figure}

To demonstrate this prediction with an example, we use a 
linear combination of Chebyshev polynomials with maximal 
degree 21 to approximate the sign function in the overlap
formula (\ref{signH}).
We consider again a typical configuration at $\beta =6$ on a 
$4^{4}$ lattice. Fig.\ \ref{sign-poly} shows the resulting
spectra if we start from the Wilson fermion resp.\ the HF, 
and we see that the latter is clearly superior.

\subsection{Approximation of the inverse square root}

We now come to a second way to approximate the overlap
formula in terms of polynomials. The idea is not to
consider the sign function any more, but to approximate
directly the inverse square root in eq.\ (\ref{overlap}). 
This was first applied in Ref.\ \cite{HJL}.
If we start from an approximate GW operator $D_{0}$, then
the square root is close to 1 (or in general close to
$\mu$), and the direct expansion around this constant looks
attractive. We have tested this method in the Schwinger
model and we found a very fast convergence if we start
from the 2d HF, but it slows down if we start from the
Wilson fermion \cite{BH}. To get started with the
polynomial approximation, one re-scales the operator $A_{0}$
so that the spectrum of $A^{\dagger}_{0}A_{0}$ is all contained
in an interval $[\delta , 1]$  \quad $(0< \delta \ll 1)$.
The virtue of the
function to be expanded, $f(x) = 1/\sqrt {x}$,
is its continuity in this interval.
However, if $\delta$ becomes very small we approach
a singularity, which is a problem similar to the
discontinuity of the sign function.
Hence the question here is how small $\delta$ is going to be.
Actually the question is the same as in Subsection 4.1
(and also the re-scaling is the same),
because $A^{\dagger}_{0}A_{0} = H^{2}$. So we can refer again
to Fig.\ \ref{histoH} where one just has to square the eigenvalues.

Also the comparison of spectra looks very similar to Fig.\
\ref{sign-poly}, so we do not repeat it but turn to a
systematic study of the convergence to an overlap fermion.

\section{Convergence rate}

We now look explicitly at the convergence of the overlap formula
approximated by polynomials. 
\footnote{The results of this and the following Sections
have been summarized before in Ref.\ \cite{lat2001}.}
As a measure for the deviation from the final overlap fermion,
we measure the maximal deviation of any
eigenvalue in the spectrum of $\frac{1}{\mu} D$ from the
unit circle with center 1, $\{ z \ \rule[-0.7mm]{0.4mm}{4mm}  \ 
\vert z-1 \vert = 1 \ \}$. For an exact
GW fermion --- with respect to the standard kernel (\ref{Rst}) ---
the spectrum is situated exactly on this circle, as we pointed out 
in Section 1.

\begin{figure}[hbt]
\def\fpsangle{270}
\epsfxsize=80mm
\fpsbox{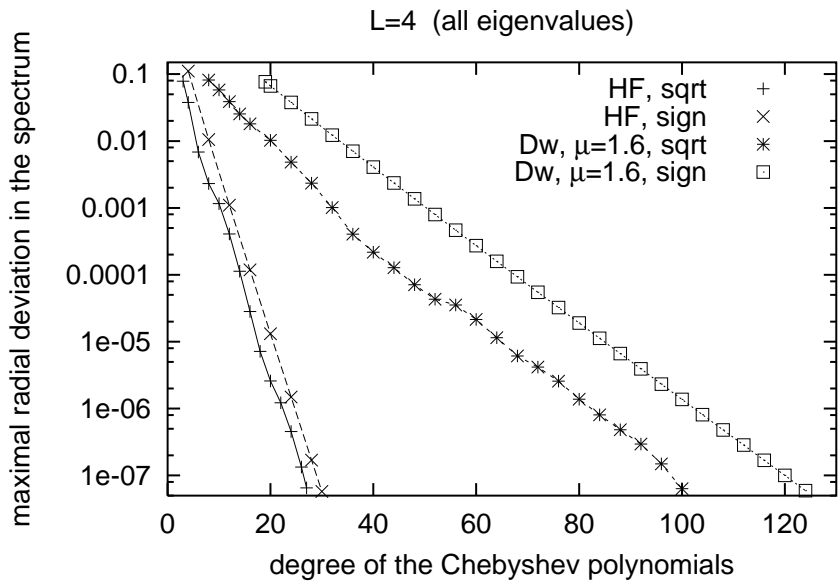}
\vspace{-3mm}
\caption{\it{The maximal radial deviation from the normalized
Ginsparg-Wilson circle, evaluated from of the full spectrum
on a $4^{4}$ lattice. We compare the HF at $\mu =1$
with the Wilson fermion at $\mu =1.6$, in both cases expanding
the inverse square root as well as the sign function.}}
\label{deviL4}
\end{figure}

Fig.\ \ref{deviL4} shows this maximal deviation for
a typical configuration on a $4^{4}$ lattice, as a
function of the polynomial degree. We consider both, the
expansion of the sign function as well as the expansion
of the inverse square root. The converge rate is clearly
{\em exponential} \, in the degree of the polynomial.
We see further that the HF converges
much faster than the Wilson fermion. For the comparison 
between the two polynomial expansions one has to take into
account that the square root expansion refers to
$H^{2}$, whereas the sign function is expanded in $H$.
Hence the former actually picks up a factor of 2 in the
comparison of the polynomial degree, which makes the sign
function look favorable, especially for the HF.

\begin{figure}[hbt]
\def\fpsangle{270}
\epsfxsize=80mm
\fpsbox{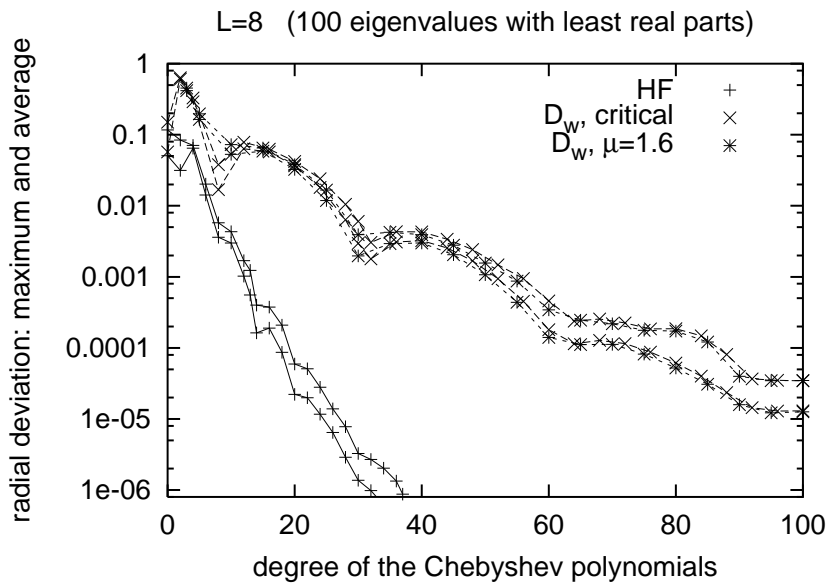}
\vspace{-3mm}
\caption{\it{The maximal radial deviation and the mean radial
deviation from the normalized
Ginsparg-Wilson circle, evaluated from of the 100 energy
eigenvalues with the lowest real parts (physical branch)
on a lattice volume $8^{4}$. We compare the inverse square root 
expansion for the HF and for the
Wilson fermion. In the latter case we also compare the case of
the free hopping parameter and a suitable mass parameter 
of $\mu =1.6$ with the critical Wilson fermion at $\mu =1$.
They behave very similarly, and the HF converges much faster.}}
\label{deviL8}
\end{figure}

Since the volume considered in Fig.\ \ref{deviL4}
is quite small, we now proceed to a $8^{4}$ lattice.
Here we cannot evaluate the full spectrum any more,
but with the Arnoldi algorithm we can identify a selected
set of eigenvalues. We thus evaluated the 100 eigenvalues
with the least real part (as shown in Fig.\ \ref{non+ultra}
before), because they are physically most relevant, 
and we measured for this subset again the maximal
radial deviation. In Fig.\ \ref{deviL8} we plot this maximal 
deviation, and also the mean deviation, again as a function
of the polynomial degree. For the Wilson fermion we consider
two options: we take either the free hopping parameter and optimize
the mass parameter to $\mu = 1.6$, or we take $\mu =1$, as in the 
case of the HF, and insert the critical Wilson fermion.
The goal is to make sure that we compare the HF really to the best 
application of the Wilson fermion. However, as Fig.\ \ref{deviL8} shows,
the difference between the different ways to use the Wilson fermion
is tiny.

Due to the arbitrary truncation at just 100 eigenvalues
the exponential behavior is not as clean as in the case of
the full spectrum. However, we see that the behavior is
very similar, and the improvement of the HF clearly persists
in the same magnitude.\\

To make these observations more quantitative, we discuss as an 
example the sign expansion on the $4^{4}$ lattice, which has
a very precise exponential behavior. For some configuration typical
at $\beta =6$ we measured for the Wilson fermion the maximal deviation
$d_{W}^{max}(n) = \exp (-0.134 n)$, where $n$ is the degree of the Chebyshev
polynomial, and for the mean deviation we found
$d_{W}^{mean}(n) = 0.13 \exp (-0.134 n)$, hence the exponential factor
is the same. For the same configuration, the following
HF deviations where obtained:
$d_{HF}^{max}(n) = \exp (-0.737 n)$, and
$d_{HF}^{mean}(n) = 0.1 \exp (-0.737 n)$.

We mention two ways how to arrive at conclusions from these
numbers. 
\begin{itemize}
\item First, we could fix some degree $n$ which we consider
affordable in a simulation. The precision of the GW
approximation compares as
\begin{equation}
\frac{d^{max}_{W}(n)}{d^{max}_{HF}(n)} \cong e^{0.6n} \ .
\end{equation}
For realistic degrees like $n= 20 \dots 100$ this ratio
of the accuracies takes a very considerable magnitude.

\item On the other hand, we could fix a certain accuracy $d^{max}$
which we consider necessary to trust the chiral quality of the
approximated GW fermion. Then the polynomial
degrees, which are required to provide this precision,
compare as
\begin{equation}
\frac{n_{W}}{n_{HF}} \cong 5.5 \ .
\end{equation}
This factor may be regarded as the effective gain of the HF due to
the faster convergence, since the computational effort
is essentially proportional to $n$.
\end{itemize}

The fluctuation of these ratios over different configurations
are modest, even though $d^{max}$ may vary significantly.
For a systematic statistical study we now move on to larger 
lattices.

\section{Condition numbers}

\begin{figure}[hbt]
\def\fpsangle{270}
\epsfxsize=57.5mm
\fpsbox{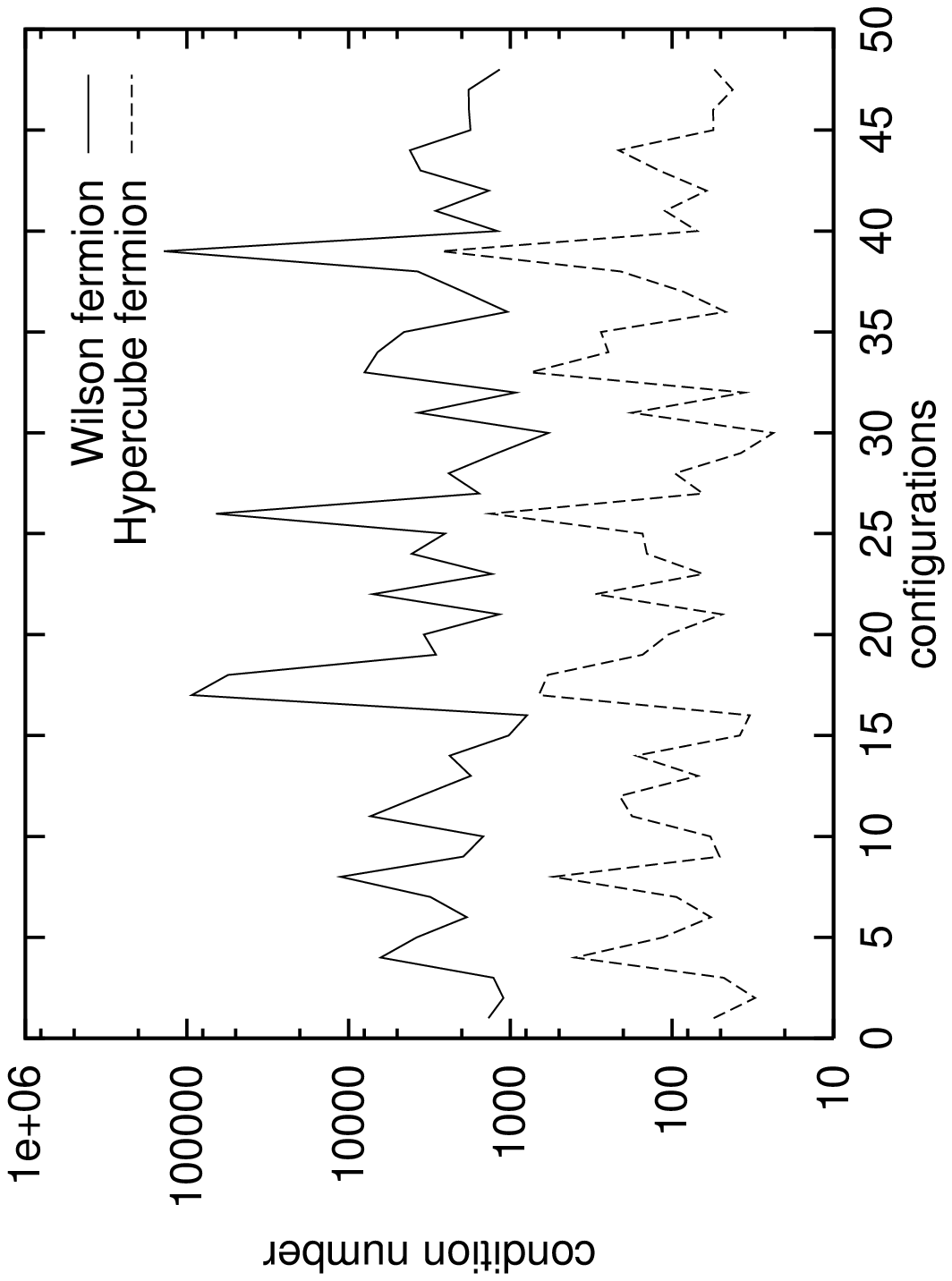}
\def\fpsangle{270}
\epsfxsize=51.5mm
\fpsbox{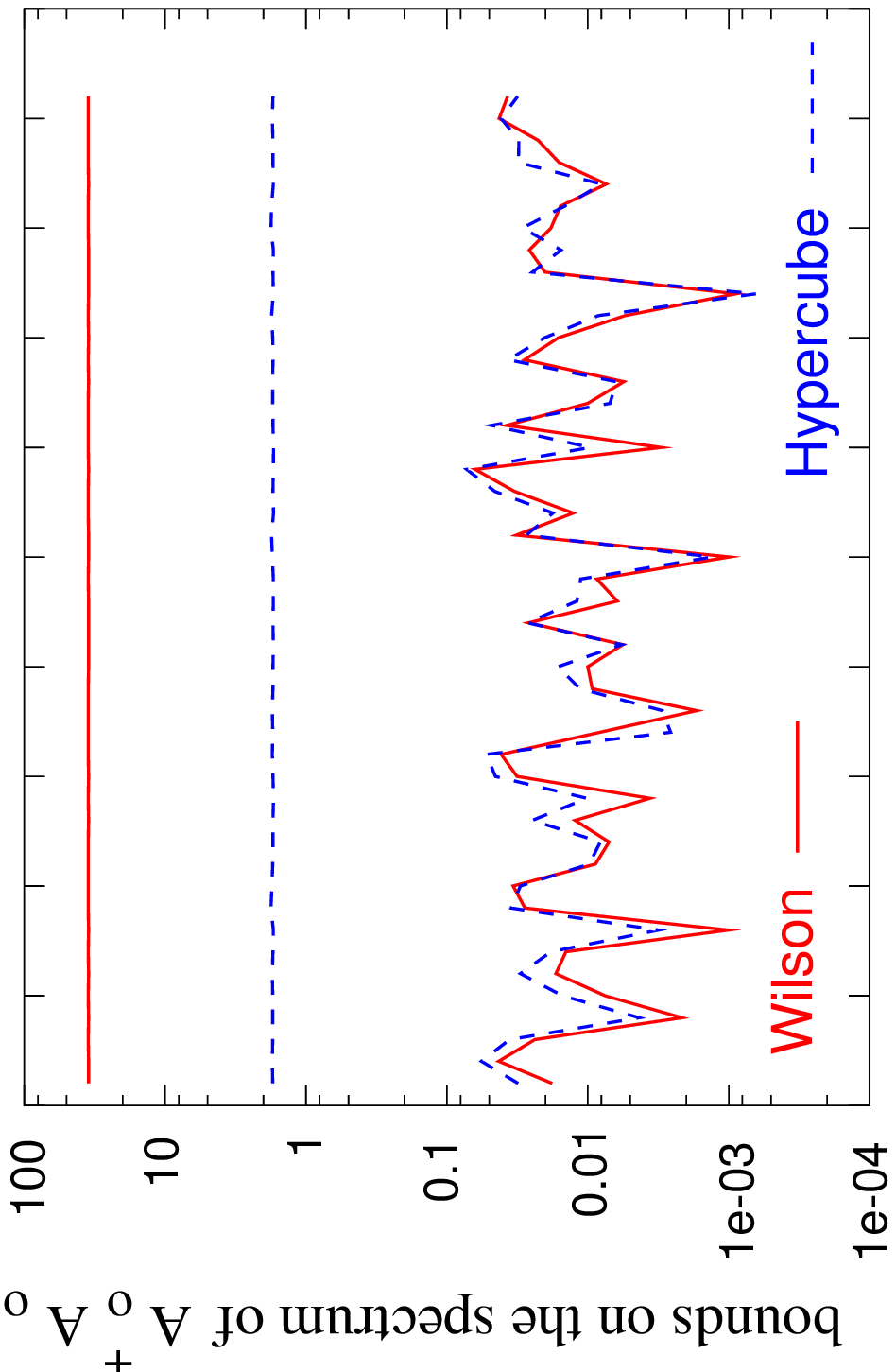}
\vspace{-3mm}
\caption{\it{On top: a history of the condition numbers
of $A_{0}^{\dagger}A_{0}$ at $\beta =6$ on a $12^{4}$ lattice.
Below: the corresponding history of the upper and lower bound 
of the spectra. It shows that the improvement of the condition
number of the HF is essentially due to the {\em upper} bound.}}
\label{cond-fig}
\vspace*{-4mm}
\end{figure}

After the explicit convergence study of the last Section,
we now turn our attention to the condition numbers of the
operators $A_{0}^{\dagger} A_{0}$, which are crucial for the
convergence rate. This allows us to proceed to larger lattices
of size $12^{4}$, still at $\beta =6$. We first show a history
for the condition numbers for the HF and the Wilson fermion in
Fig.\ \ref{cond-fig} (on top). 
Here we adapted the parameters so that they are 
optimal on the larger lattice. For the HF the new set of 
optimal parameters reads
\begin{equation}
u =0.75, \quad v=0.99, \quad
\alpha = 0.87 , \quad c = 0.01 , \quad \mu =1.3 \ .
\end{equation}
For the Wilson fermion (with the free hopping parameter), 
$\mu = 1.64$ turned out to be optimal with respect to the
condition number, though the dependence on $\mu$ is weak ---
as it is also the case for locality, see Section 6. Hence the 
condition numbers at $\mu =1.4$ --- which is optimal with respect
to locality \cite{HJL} --- 
differ only little from the result in Fig.\ \ref{cond-fig}.

We see that the HF condition number is improved typically
by one to two orders of magnitude. Since it
is defined by the ratio of the upper bound divided by the lower bound,
it is instructive to consider these bounds separately. Their
histories are shown in Fig.\ \ref{cond-fig} (below). They reveal that 
the improvement of the condition number is almost entirely an effect
due to the {\em upper} bound.

\begin{figure}[hbt]
\def\fpsangle{270}
\epsfxsize=80mm
\fpsbox{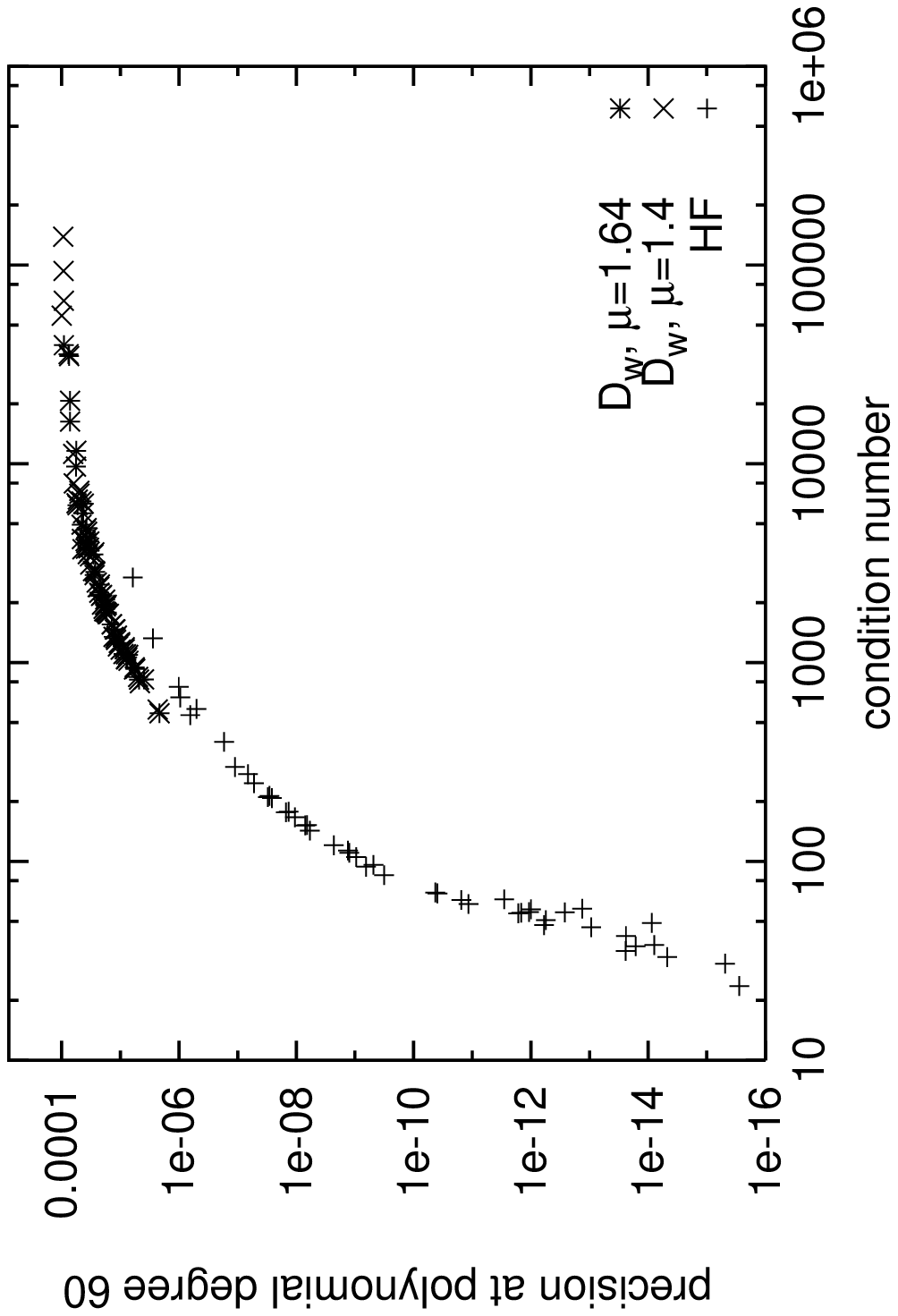}
\vspace{-3mm}
\caption{\it{The precision of the overlap formula approximated
by Chebyshev polynomials of the moderate degree of 60.
As a measure for the precision, we show the accuracy of the
function $f_{\rm max}(r=24)$ (c.f.\ Section 6) for a set of 
configurations at $L=12$, $\beta =6$.
We plot this accuracy against the condition number, in order
to illustrate their monotonous relation, and the progress of the
HF over the Wilson fermion.}}
\vspace*{-3mm}
\label{cond-conv}
\end{figure}

In Fig.\ \ref{cond-conv} we illustrate explicitly how the condition 
number translates into an accelerated convergence. As a (somewhat
arbitrary) measure for the speed of convergence, we consider
Chebyshev polynomials at some moderate degree, $n=60$, and
measure the deviation of the function $f_{\rm max}(r=24)$ from the exact
result (obtained from huge values of $n$). (The quantity 
$f_{\rm max}(r=24)$ represents the maximal correlation over the
largest distance on our lattice;
an explicit definition will be given in Section 6.)
A polynomial of degree 60
may be affordable in simulations, and we see that it
typically approximates $f_{\rm max}(r=24)$ already to a high accuracy for
the HF, but not for the Wilson fermion. For the latter we see
that $\mu =1.64$ is better for the condition number than the
locality optimal mass parameter of $\mu =1.4$.\\

However, it is important to note that most practical applications
of overlap fermions are performed such that the lowest few modes
are projected out and treated separately. Then the above
polynomial evaluation concerns the rest of the spectrum.
This method helps a lot, because often very few modes are
responsible for a slow convergence. However, their separate
treatment is also tedious, because it requires a very accurate
determination of the corresponding eigenfunctions.
Hence the number of modes to be projected out is usually not
more than about 15; beyond that the spectrum becomes quite dense,
so projecting out further eigenvalues raises the remaining
lower bound only very little.

\begin{figure}[hbt]
\def\fpsangle{270}
\epsfxsize=80mm
\fpsbox{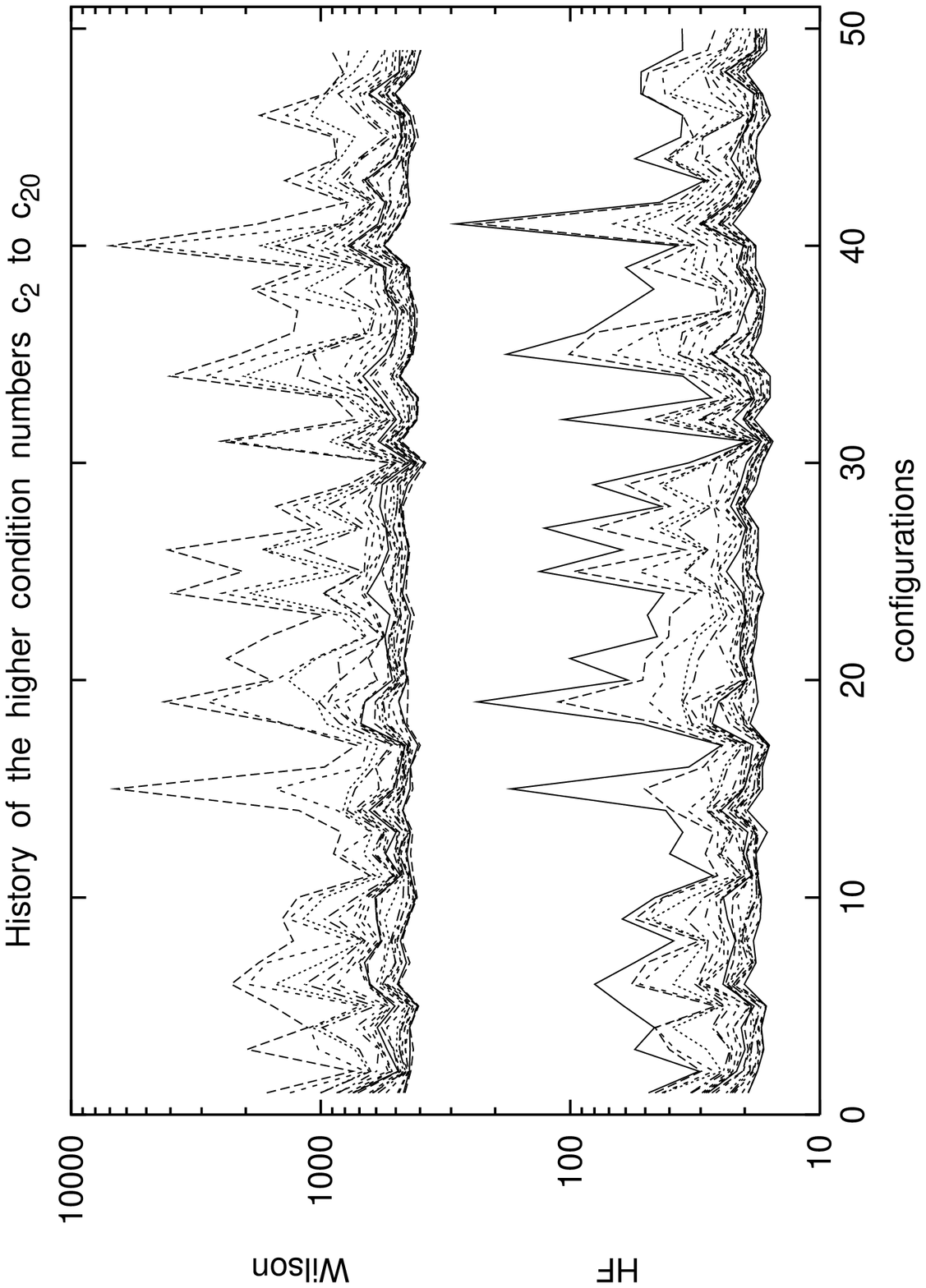}
\vspace{-3mm}
\caption{\it{The history of the ``higher condition numbers''
$c_{2} \dots c_{20}$ (defined in eq.\ (\ref{highcon}))
for the operator $A_{0}^{\dagger}A_{0}$,
built from the HF (at $\mu =1.3$) and from the 
Wilson fermion (at $\mu =1.64$).}}
\label{histcond_c2-20}
\vspace*{-3mm}
\end{figure}

\begin{figure}[hbt]
\def\fpsangle{270}
\epsfxsize=80mm
\fpsbox{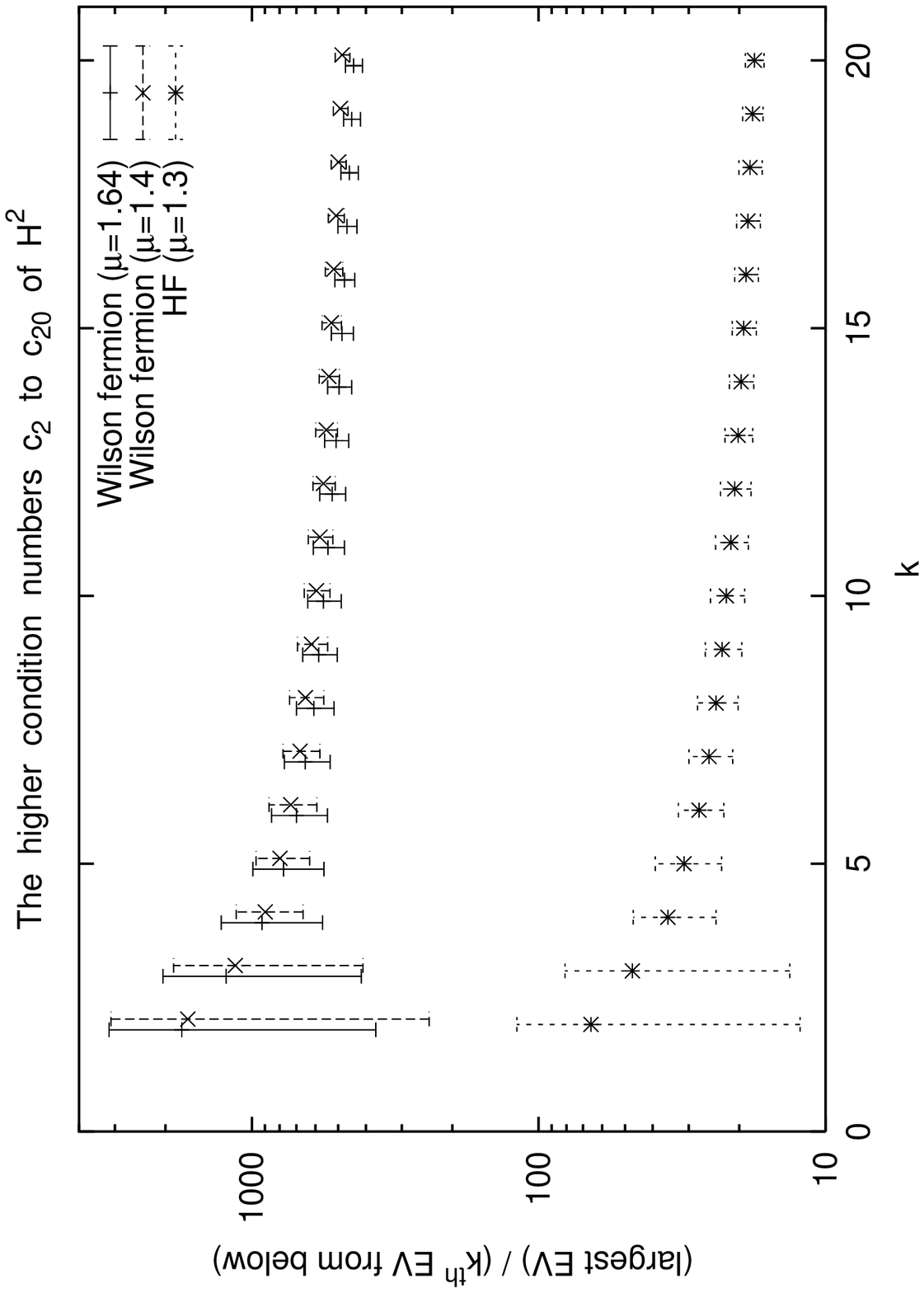}
\vspace{-3mm}
\caption{\it{The expectation values for the higher condition numbers
$c_{k}$ ($k=2 \dots 20$) (defined in eq.\ (\ref{highcon}))
for the operator $A_{0}^{\dagger}A_{0}$,
built from the HF and from the Wilson fermion.
In the most popular regime (around $k=15$) the HF gains a factor
of $\approx 25$ over the Wilson fermion. For the latter we also
confirm that $\mu =1.64$ is a little better for the condition number
than the locality optimal parameter $\mu=1.4$.}}
\label{cond_c2-20}
\vspace*{-4mm}
\end{figure}

To do justice to this situation, we introduce ``higher condition
numbers'' $c_{k}$, which are defined by the ratio 
\begin{equation}  \label{highcon}
c_{k} = \frac{{\rm largest~eigenvalue}}
{k^{\rm th} {\rm ~eigenvalue~from~below}} \ .
\end{equation}
If one projects out 15 eigenvalues, for example, then $c_{16}$
is relevant for the convergence of the polynomial evaluation.
In Fig.\ \ref{histcond_c2-20} we show the histories 
for the higher conditions numbers
$c_{2} \dots c_{20}$, for the HF and the Wilson fermion.
This plot confirms that around $k=15$ the eigenvalue density
is large already, and beyond they become even more dense.
Hence it is hardly motivated to increase
$k$ much further. We also see that
these histories are much smoother compared to $c_{1}$
(which was shown in Fig.\ \ref{cond-fig}),
so here we are able to take sensible mean values.
The results are shown in Fig.\ \ref{cond_c2-20}, and 
we see that in the relevant regime the improvement factor for the 
HF amounts to about $25$. The convergence rate
behaves like the square root of the relevant condition number.
\footnote{Note that the relevant condition number
corresponds to $1/\delta^{2}$ in the notation of Section 3.}
Hence we gain a factor of $\approx 5$. This is amazingly
consistent with the explicit result obtained on the small 
lattices (but from the full spectrum) in Section 4.

\section{Locality}

Since there is occasionally some confusion about the term
``locality'' of a lattice action, 
we first clarify that we refer to the definition
that all correlators decay at least exponentially
in the distance. This is the property which is crucial
for providing a safe continuum limit (since the decay width
is fixed in lattice units).

It was conjectured \cite{EPJC} --- and later proved \cite{ultra}
--- that Ginsparg-Wilson fermions cannot be ``ultralocal'', 
not even in the free case.
This means that their couplings cannot drop to zero beyond a 
finite number of lattice spacings. However, locality in the
above sense was shown for the free perfect fermion \cite{QuaGlu}
and for the Neuberger fermion \cite{ML} to hold. In Ref.\ \cite{EPJC}
it was shown that the truncated perfect free HF leads to 
an overlap-HF, which is more local than the Neuberger fermion
(it has a faster exponential decay).

In the presence of gauge interactions, locality can be proved
for smooth configurations, either by assuming a small upper
limit for the deviation of any plaquette variable from 1
\cite{HJL,Neu2}, or by assuming that the eigenvalues of 
$A^{\dagger}_{0} A_{0}$ 
do not cluster densely in the vicinity of zero \cite{HJL}.
For realistic configurations, the exponential decay was observed
statistically for the Neuberger fermion at $\beta =6$ \cite{HJL}.
This property may collapse at much stronger coupling, but at some 
point the overlap formula is not applicable any more also for other
reasons, as we discussed in Section 2. The statistical demonstration
was done by showing that the ``maximal correlation'' between any
two lattice sites, separated by a taxi driver distance $r$,
decays exponentially in $r$. More precisely, the expectation value
of the function
\begin{equation}  \label{fmax}
f_{\rm max}(r) :=  \ ^{\rm max}_{x,\, y} \ \Big\{ \ \Vert \psi (y) \Vert 
\ \rule[-0.7mm]{0.4mm}{4mm}  \ 
\sum_{\mu =1}^{4} \vert x_{\mu}-y_{\mu} \vert = r \ \Big\}
\end{equation}
has to decay exponentially, if a unit source is located at the
(arbitrary) site $x$. We probed 6 sites $x$ for each configuration,
then we went through all sites $y$ to determine the maxima.
The 6 options for $x$ were sufficient to stabilize the function 
$f_{\rm max}(r)$. 
(Remember that this function was mentioned before in Section 5.)

\begin{figure}[hbt]
\hspace*{0.5cm}
\def\fpsangle{270}
\epsfxsize=80mm
\fpsbox{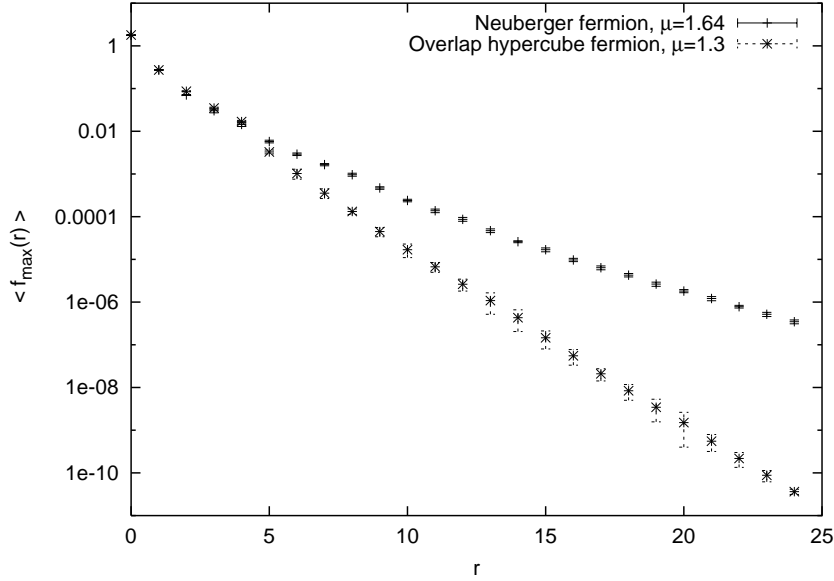}
\vspace{-3mm}
\caption{\it{Comparison of the degree of locality for the 
Neuberger fermion and for the overlap HF. We see that the
latter is clearly more local; the exponent differs by more than
a factor of 2.}}
\label{fig-loc}
\vspace*{-3mm}
\end{figure}

As we explained in the introduction, the property
(\ref{approx}) suggests that the higher degree of locality
of the overlap-HF may also persist in the presence of gauge
interactions. In fact, in the Schwinger model a comparison
of the function $\langle f_{\rm max}(r) \rangle$ confirmed this
conjecture \cite{BH}. 
Here we extend this comparison to
QCD on a periodic $12^{4}$ lattice, which is the size that was 
also used in Ref.\ \cite{HJL}. The use of the taxi driver 
metrics allows us to proceed to maximal distance 24,
and the exponential decay is clearly visible.
\footnote{Of course, the decay is also exponential
with respect to the Euclidean distance.
There the decay is less smooth, but the asymptotic behavior
can be extracted as $0.04 \exp(-\vert x - y \vert)$ for the
Neuberger fermion, and as $0.7 \exp(-2 \cdot 
\vert x - y \vert)$ for the overlap HF.}
Our result is shown in Fig.\ \ref{fig-loc}.
For the Neuberger fermion it agrees well
with the result of Ref.\ \cite{HJL}
(which was also obtained at $\beta = 6$),
although we used a different mass parameter.
We inserted $\mu =1.64$ which was optimal
for the condition number of the Neuberger fermion
(c.f.\ Section 5), whereas
Ref.\ \cite{HJL} used $\mu =1.4$, which is slightly better
for the locality --- because it was optimized with this respect ---
but the difference is really small. 
Also for the overlap-HF we used the parameter which
is optimal for the condition number, in that case $\mu =1.3$.

We see that the overlap-HF is by far more local. To be explicit, 
the asymptotic decay of $\langle f_{\rm max}(r) \rangle$ behaves like
$0.017 \exp (-0.45 r)$ for the Wilson fermion, and like
$0.017 \exp (-0.93 r)$ for the HF.
This observation suggests that the range of applicability of
the overlap formula extends up to stronger coupling for
the overlap-HF.

Interestingly, the very clean exponential decay sets in
only in the presence of interactions. For free
fermions, the decay is faster, of course, but the exponential
behavior is not as exact as in Fig.\ \ref{fig-loc}.

On the technical side we mention that the convergence
of $f_{\rm max}(r)$ at long distances requires a very precise
evaluation of the inverse square root. This motivated the use
of $f_{\rm max}(r=24)$ as a measure for the precision of the
approximations to the overlap formula, see Section 5.

\section{Approximate rotation invariance}

\begin{figure}[hbt]
\hspace*{0.5cm}
\def\fpsangle{270}
\epsfxsize=80mm
\fpsbox{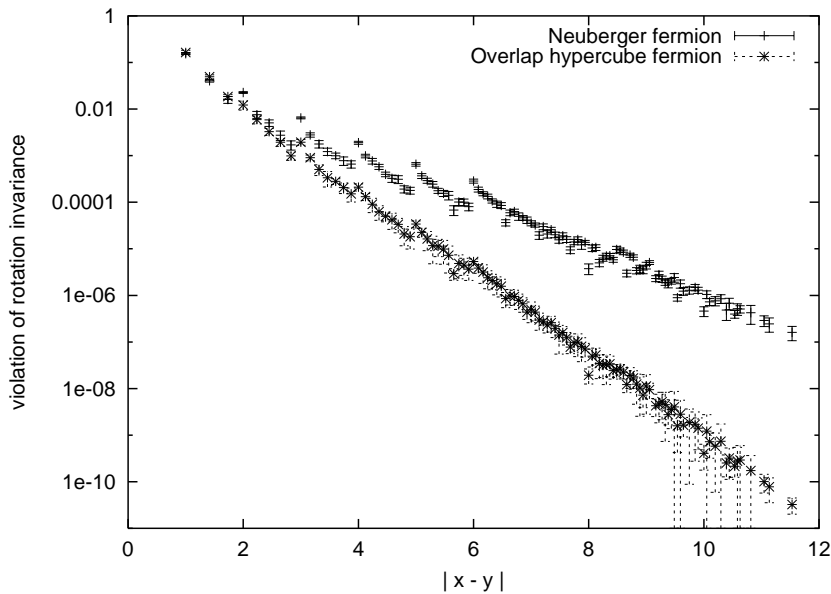}
\vspace{-3mm}
\caption{\it{Comparison of the violations of the rotation
invariance for the Neuberger fermion and for the overlap-HF. 
We see that the latter approximates rotation invariance much
better. In both cases, the violations of rotation invariance
decay exponentially with the distance, but for the overlap
HF the exponent of the decay is almost doubled.}}
\label{fig-rot}
\vspace*{-3mm}
\end{figure}

As a last comparison between the Neuberger fermion and the 
overlap HF, we consider the extent of the violations of
rotation invariance. To quantify this property, we introduce
the function $f_{\rm max}(\vert x - y \vert )$, which corresponds
to definition (\ref{fmax}) in Euclidean metrics, as well as
the corresponding function $f_{\rm min}(\vert x - y \vert )$,

\begin{eqnarray}
f_{\rm max}(\vert x-y \vert ) & := &  
\ ^{\rm max}_{x,\, y} \ \Big\{ \ \Vert \psi (y) \Vert 
\ \rule[-0.7mm]{0.4mm}{4mm}  \ 
\Big[ \sum_{\mu =1}^{4} (x_{\mu}-y_{\mu})^{2} \Big]^{1/2} 
= \vert x-y \vert \ \Big\} \ , \nonumber \\
f_{\rm min}(\vert x-y \vert ) & := &  
\ ^{\rm min}_{x,\, y} \ \Big\{ \ \Vert \psi (y) \Vert 
\ \rule[-0.7mm]{0.4mm}{4mm}  \ 
\Big[ \sum_{\mu =1}^{4} (x_{\mu}-y_{\mu})^{2} \Big]^{1/2} 
= \vert x-y \vert \ \Big\} \ .
\end{eqnarray}
We put a unit source on one site $x$ and probe all other sites $y$ for
a fixed Euclidean distance. For each distance $\vert x - y \vert$
that occurs, we determine the difference
\begin{displaymath}
\langle f_{\rm max}(\vert x-y \vert ) - f_{\rm min}(\vert x-y \vert )
\rangle
\end{displaymath}
which represents a measure for the violation of rotation
invariance. Fig.\ \ref{fig-rot} shows that this difference decays
exponentially as a function of the Euclidean distance.
(Again we present data from a periodic $12^{4}$ lattice at $\beta =6$.)
For the Neuberger fermion (at $\mu = 1.64$), 
the asymptotic decay amounts to $0.065 \exp(-0.11 \cdot \vert x-y \vert )$.
The corresponding asymptotic decay for the overlap HF (at $\mu =1.3$)
behaves as $\exp(-2.10 \cdot \vert x - y \vert )$.

\section{Conclusions}

We have constructed a HF for QCD, which approximates
the GWR at $\beta =6$. It involves the couplings of the truncated
perfect free fermion, along with four more parameters which
amplify certain links and add fat links as well as a clover term.
This HF is designed to be inserted into the overlap formula,
which renders its chirality exact. Since we start off in the right
vicinity, the convergence under polynomial evaluations of the
overlap formula is accelerated compared to the standard Neuberger
fermion, which uses the Wilson fermion as point of
departure. We observed consistently over lattice volumes
$4^{4}$ up to $12^{4}$ an improved convergence rate the overlap-HF,
which allows us to obtain the same chiral accuracy as the Neuberger
fermion if the polynomial degree is reduced by a factor 
$\gsim 5$. This implies a correspondingly reduced computational
effort.

Our experience with the numeric treatment of the HF is based on
a simple implementation \cite{precon}, which first computes
(hierarchically)
a look-up table for all the ``hyperlink variables'' across 
(spatial) diagonals inside the unit hypercube. This increases
the number of $SU(3)$ matrices per site to be stored by a factor
of 10 compared to the clover Wilson fermion.
We further estimate that the computational 
overhead amounts to a factor $\approx 15$.
Hence after the gain in the convergence rate we are
left with an computational overhead of a factor of $\approx 3$.

However, we should emphasize that the algorithms for the
overlap-HF are not optimized carefully yet, so there may still 
be space for a further gain due to a better implementation.
Moreover, we expect a number of further virtues
of the overlap-HF compared to the Neuberger fermion,
which could well compensate this remaining overhead.
We have shown that locality is clearly improved ---
the exponent of the correlation decay (in the distance 
between $\bar \psi $ and $\psi$) is doubled --- which
makes the overlap formula applicable up to stronger coupling.
We have also shown that the overlap-HF approximates
the rotation invariance very well, in contrast to the Neuberger
fermion.

However, the really crucial question is if our expectation for 
an improved scaling can also be confirmed. If the overlap-HF
allows for the use of a somewhat larger physical lattice 
spacing, then the remaining overhead could be more than compensated.
This hope is based in particular on the elements of
an truncated perfect actions, which are incorporated
in our HF construction. Indeed, a strongly improved scaling was
observed for free fermions and in the 2-flavor Schwinger model
$d=2$ \cite{BH}. We also run some
tests for the meson dispersions in QCD, but it turned out
that the dispersions are quite noisy, in particular for the 
overlap-HF. Hence we postpone the delicate question of scaling 
for further investigation. Interestingly, the phenomenon of more 
noise is also known from $O(a)$ improved Wilson fermions
with a clover term. As a general property, GW fermions are also 
$O(a)$ improved. Hence it is conceivable that the overlap HF
has an even better scaling behavior than the original HF,
since the overlap formula removes the $O(a)$ artifacts.

As long as GW fermions can only be applied in the quenched
approximation, the connection with chiral perturbation
theory \cite{DDHJ} seems most attractive. However,
the ultimate goal is the use of dynamical Ginsparg-Wilson 
fermions, which appears as a tremendous task. Hence it
is worthwhile studying the optimal access carefully.
We hope that the current work contributes to this optimization.\\

{\it {\bf Acknowledgment} \ \ 
I am very much indebted to Ivan Hip, who made important
contributions to this work. I also thank him, as well as 
David Adams, Norbert Eicker, 
Philippe de Forcrand, Leonardo Giusti,
Karl Jansen, Waseem Kamleh, Thomas Lippert, 
Martin L\"{u}scher, Klaus Schilling, Rainer Sommer 
and Urs Wenger for useful 
discussions. The computations for this work were performed
on the NICSE machines at the Forschungszentrum J\"{u}lich.}

\end{document}